\documentclass[fleqn,usenatbib]{mnras}

\usepackage{newtxtext,newtxmath}

\usepackage[utf8]{inputenc}

\usepackage[T1]{fontenc}

\defcitealias{smi04}{Smith 2004}
\defcitealias{smi18}{Smith et al. 2018}
\defcitealias{bor19}{Bordiu \& Rizzo 2019}
\defcitealias{rem13}{Remmer et al.  2013}
\defcitealias{abr14}{Paper~I}
\defcitealias{hum99}{Humphreys et al. 1999}
\defcitealias{dor04}{Dorland et al. 2004}
\defcitealias{dav97}{Davidson et al. 1997}
\defcitealias{art11}{Artugau et al. 2011}
\defcitealias{ver05}{Verner et al.  2005}
\defcitealias{meh10}{Mehner et al. 2010}
\defcitealias{paperII}{Paper II}
\defcitealias{abr20}{Abraham et al. 2020}
\usepackage{graphicx}	

\title[Telluric absorption lines $\eta$ Car]{Telluric absorption lines in the ALMA spectra of $\eta$ Car}

\author[Abraham et al.]{
Zulema Abraham,$^{1}$\thanks{E-mail: zulema.abraham@iag.usp.br}
Pedro. P. B. Beaklini,$^{2}$
Pierre Cox,$^{3}$
Diego Falceta-Gon\c calves$^{4}$
\newauthor
and 
Lars-\AA ke  Nyman$^{5}$
\\
$^{1}$Instituto de Astronomia, Geof\'isica e Ci\^encias Atmosf\'ericas, Universidade de S\~ao Paulo \\
Rua do Mat\~ao 1226, CEP 05508-090, S\~ao Paulo, Brazil\\
$^{2}$National Radio Astronomy Observatory, 1003 Lopezville Road, Socorro, NM 87801, USA\\
$^{3}$Institut d'Astrophysique de Paris, Sorbonne Universit\'e, UPMC Universit\'e Paris 6 and CNRS,\\ UMR 7095, 98bis boulevard Arago, 75014, Paris, France\\
$^{4}$Escola de Artes, Ci\^encias e Humanidades, Universidade de S\~ao Paulo,\\ Rua Arlindo Bettio 1000, CEP 03828-000, S\~ao Paulo, Brazil\\
$^{5}$European Southern Observatory, Alonso de C\'ordoba 3107, Vitacura, Chile}

\date{Accepted XXX. Received YYY; in original form ZZZ}

\pubyear{2020}

\begin{document}
\label{firstpage}
\pagerange{\pageref{firstpage}--\pageref{lastpage}}
\maketitle

\begin{abstract}
The massive binary system formed by $\eta$ Car and an unknown companion is a strong  source at millimetre and submillimetre wavelengths. Close to the stars, continuum bremsstrahlung and  radio recombination lines originate in the massive ionized wind of $\eta$ Car and in several compact sources of high density plasma. Molecular lines are also detected at these wavelengths, some of them are seen in absorption towards the continuum emission region,  many of them  revealed by ALMA observations. However, because the ALMA atmospheric calibration is performed in a low spectral resolution mode, telluric lines can still be present in some high-resolution spectra of scientific products, which could lead to a false identification of molecules. 
In this work, we  explore three different sets of ALMA archive data of $\eta$ Car, including  high resolution ($0\farcs065\times 0\farcs043$) observations recently published by our group, to verify which of these absorption lines are real and discuss their origin. We conclude that some of them truly originate  in clouds close to the binary system, while others are artifacts of a faulty elimination of telluric  lines during ALMA calibration procedure. We found that these absorption lines are not present in the phase calibrators because they are much weaker than $\eta$ Car,  where the absorption line  appears because the high intensity continuum enhances the small individual systematic calibration errors.
\end{abstract}



\section{Introduction}
\label{sec:Introduction}

 $\eta$ Carinae (hereafter $\eta$ Car) is one of the most massive and luminous stars of our galaxy ($  M \sim 120$~ ${\rm M}_\odot$, $L \sim 5\times10^6$~${\rm L}_\odot$).
 It is part of a binary system in a highly eccentric  orbit; the companion star is not directly observed, but its existence was first reported by \citet{dam96} from the detection of a  5.54 y periodicity  in the light curve of the high excitation line \ion{He}{I} ${\rm \lambda}$10830.
 The periodicity was also found at at other frequencies, from radio to X and $\gamma$-rays \citep{dam00,dun03,whi04,abr05,meh10,rei15,teo16,cor17}.
 The high energy X-ray emission that varies with orbital phase reveals the existence of a shocked region formed by wind-wind collision \citep{cor01}. 
 
Although it is not possible to observe the stars directly because they are embedded in the dusty reflection nebula, the Homunculus, formed by matter ejected in the 19th century "Great Eruption", their main stellar parameters  are reasonably well established. The bolometric luminosity of $\eta$ Car, $5\times10^6$ L$_\odot$, was inferred from the infrared emission of the Homunculus nebula, assuming a distance of 2.3 kpc \citep{dah97}. Using this bolometric luminosity, a lower limit of 120 M$_\odot$ was derived for its mass from the Eddington limit,  and an  effective temperature  $T_{\rm{eff}}\sim 30,000$ K, mass loss rate  $\dot{M} \sim 10^{-3}~ M_\odot~ \rm{yr}^{-1}$  and  terminal wind velocity $v \sim 500$ km s$^{-1}$ was obtained from  NLTE models of its spectra \citep{hil01,hil06,gro12}.

The  surface temperature of the companion star,  $T\sim 37,000-40,000$~K \citep{ver05,meh10},  was inferred from the intensity of the high excitation lines observed in the spectra of the Weigelt blobs \citep{wei86}, located at e distance of $\sim 0\farcs3$ from the binary system, which cannot be excited by $\eta$ Car itself, because all the  ionizing photons are absorbed by its own massive wind \citep{abr20}. The mass of the companion star must be in the range of $30-40$ M$_\odot$, depending on its evolutionary state: main sequence, supergiant or Wolf-Rayet. A mass loss rate of $\dot M \sim 2\times 10^{-5}$ M$_\odot$ y$^{-1}$ and wind velocity  $v\sim3000$ km s$^{-1}$ were derived  fitting hydrodynamical models to a $Chandra$ X-ray grating spectrum \citep{pit02}.

Close to the binary system, dust can be formed in the wind-wind collision region \citep{fal05}. Numerical simulations show the formation of a complex and dense structure that evolves and persists during several orbital periods \citep{par11,mad12,mad13,cle14,rus16}.
While \citet{rus87}, \citet{smi95} and \citet{mor17} reported decrease in the dust emission of the Homunculus Nebula,  \citet{men19} showed that its brightness remained constant since 1968, with fluctuations of about 25\%. The study of \citet{men19} included mid-IR observations ($8-20 \mu{\rm m})$ of the Homunculus  with high spatial resolution ($0\farcs22$)  that confirmed the existence of a $5\arcsec$ diameter dusty torus and also revealed the existence of a smaller inner circumstellar disk of dust.

The binary system has a rich molecular environment, with many molecules already revealed by ALMA observations, including several transitions of the CO molecule, HCN, H$^{13}$CN and HCO$^+$. Since many molecules were detected at different wavelengths, it was not surprising that we started to see molecular absorption at ALMA frequencies \citep{smi18,bor19,mor20}. However, a recent ALMA report, \citep{hun18} showed that sometimes telluric lines are not entirely removed from the final spectrum, appearing as absorption features. Therefore, some additional analysis is required to recognize if an absorption feature is intrinsic to the source and not an artificial feature due to a calibration problem. 

\citet{hun18} explained that telluric lines like ozone could still be present as a feature in the science spectrum due to an unperfect correction of the line emission during the atmospheric calibration. Since the atmospheric calibration is performed in a low-resolution spectral mode using 128 channels of 15.625 MHz, the core emission of a few atmospheric lines is eventually not resolved. Most of those lines come from  ozone, but other telluric lines could also be present, like CH$_2$Cl$_2$. A complete list of atmospheric lines in the sub-mm range can be found in \citet{sme15}.
The lines appear when the scientific goal uses a spectral resolution higher than that used for the atmospheric calibration. It should not be a  problem if the bandpass calibrator is close in elevation to the science target, because any remaining telluric line would be corrected by the bandpass calibration. However, if the target source is  bright, even a minor elevation difference between the calibrator and the science target observation could be enough to show a remaining absorption line. We do not expect such conditions to happen often, but unfortunately, we will show that the above criteria are satisfied in some of the $\eta$~Car data.

In the paper, we  revisit some of the recent absorption lines detected  by ALMA in $\eta$~Car to check which are real and which are due to an imperfect atmospheric calibration. We included in our investigation our recent ALMA data directed to the detection of H and He recombination lines published in \citet{abr20}, where we  also detected  absorption features. None of those absorption features were discussed before, and they are not related to our conclusions based on the recombination line emission.


 \begin{figure*}
\begin{center}
\includegraphics[width=17cm]{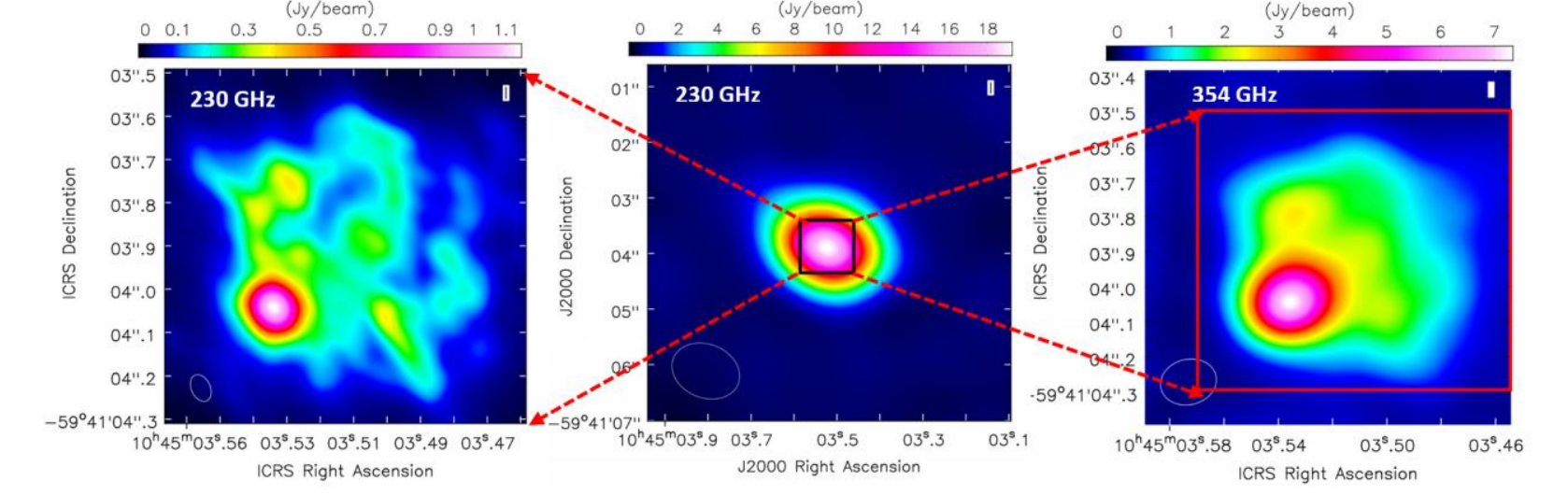}
\caption{Left panel: 230 GHz continuum raster image of $\eta$ Car, observed on  2017 November, with resolution ($0\farcs065\times 0\farcs043$). Center panel: 230 GHz continuum raster image of $\eta$ Car, reprocessed  from data extracted from ALMA public archive, observed on  2015 April, with $(1\farcs3 \times 0\farcs99)$ resolution. Right panel 354 GHz raster image, reprocessed  from data extracted from ALMA public archive, observed on  2016 October, with resolution $(0\farcs13 \times 0\farcs11)$. The scales of the continuum flux density and the synthesized beams (as white contours) are displayed on the top  and  bottom left corner of each panel, respectively.} 
\label{fig:Fig_1_telluric.pdf}
\end{center}
\end{figure*}
\begin{figure*}
\begin{center}
\includegraphics[width= 17cm]{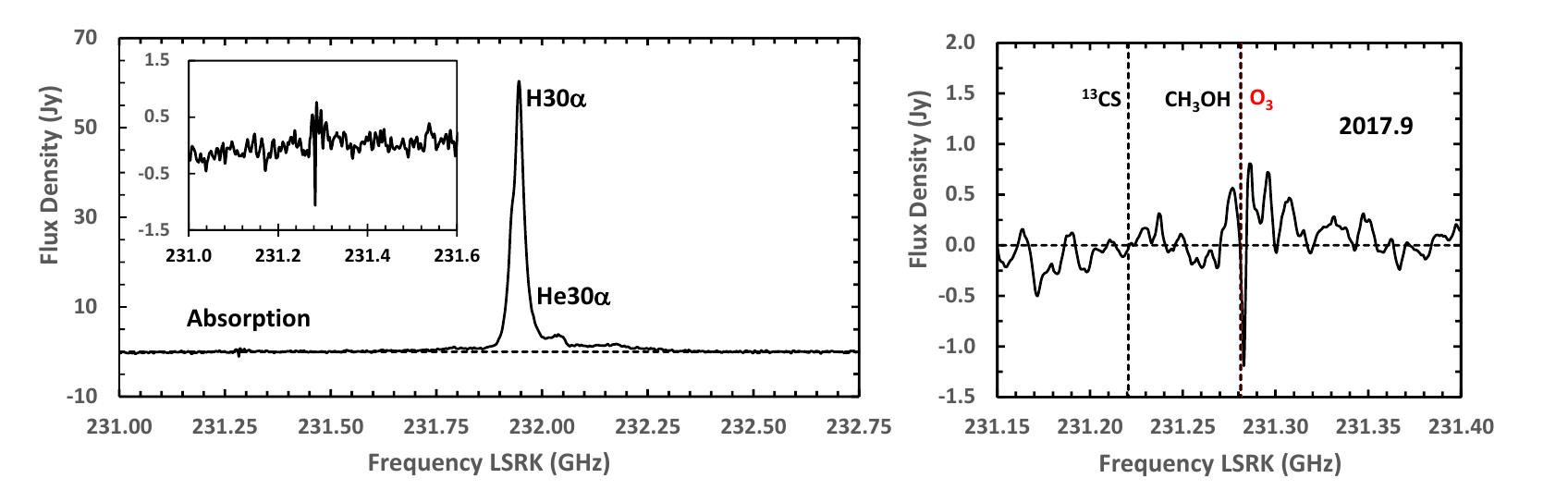}
\caption{Left panel: spectrum of the H30$\alpha$ and He30$\alpha$ emission lines observed with ALMA on 2017.9 (2017 November) \citep{abr20} in which an absorption line at the LSRK frequency of 231.28 GHz is detected (shown in an expanded scale in the insert at the top left corner). Right panel: close up of the spectrum around the absorption line. The vertical broken lines represents the frequency of the $^{13}$CS~$(5-4)$ and H$_3$COH $(10_2-9_3){\rm A^-}$  (black), and O$_3$ $[16(1,15)-16(0,16)]$ (red) transitions. }
\label{fig:Fig_2_telluric.pdf}
\end{center}
\end{figure*}

In Section \ref{sec:Observations} of this paper we describe the observations we have explored; in Section \ref{sec:Results} we present the data; in Section \ref{Discussion} we  analyze and discuss the  results and, in Section \ref{sec:Conclusions} we present the conclusions of the study.

The coordinates of  $\eta$ Car were obtained from the ICRS catalog: $\alpha(\rm J2000) =$ 10:45:3.5362, $\delta(\rm J2000) = -59$:41:4.0534; its distance is assumed to be 2300 pc \citep{dav97}, so that 0.1 arcsec corresponds to 224 AU. 


\section{Observations}
\label{sec:Observations}

We have analyzed three different sets of ALMA observations of $\eta$~Car. We included the data of our recent study \citep{abr20}, focusing on the detection of the 230 GHz continuum and the H30$\alpha$ recombination line  (project 2017.1.00725.S), because the spectral window used to measure the  recombination line has a bandwidth large enough  to include the absorption features at the frequency of 231.28 GHz reported by \citet{mor20}. Our observations include 4 spectral windows, one of them with  high-frequency resolution (948~kHz channel spacing), centered at 231.9 GHz with a bandwidth of 1.875 GHz. The other low resolution spectral windows  were centered at 230.519, 218.018, and 215.518~GHz , with similar bandwidth and channel spacing around 15~MHz, and were used to image the continuum emission. The bandpass and phase calibrators were J0904-5735 and  J1032-5917, respectively.

\begin{figure}
\begin{center}
\includegraphics[width=8cm]{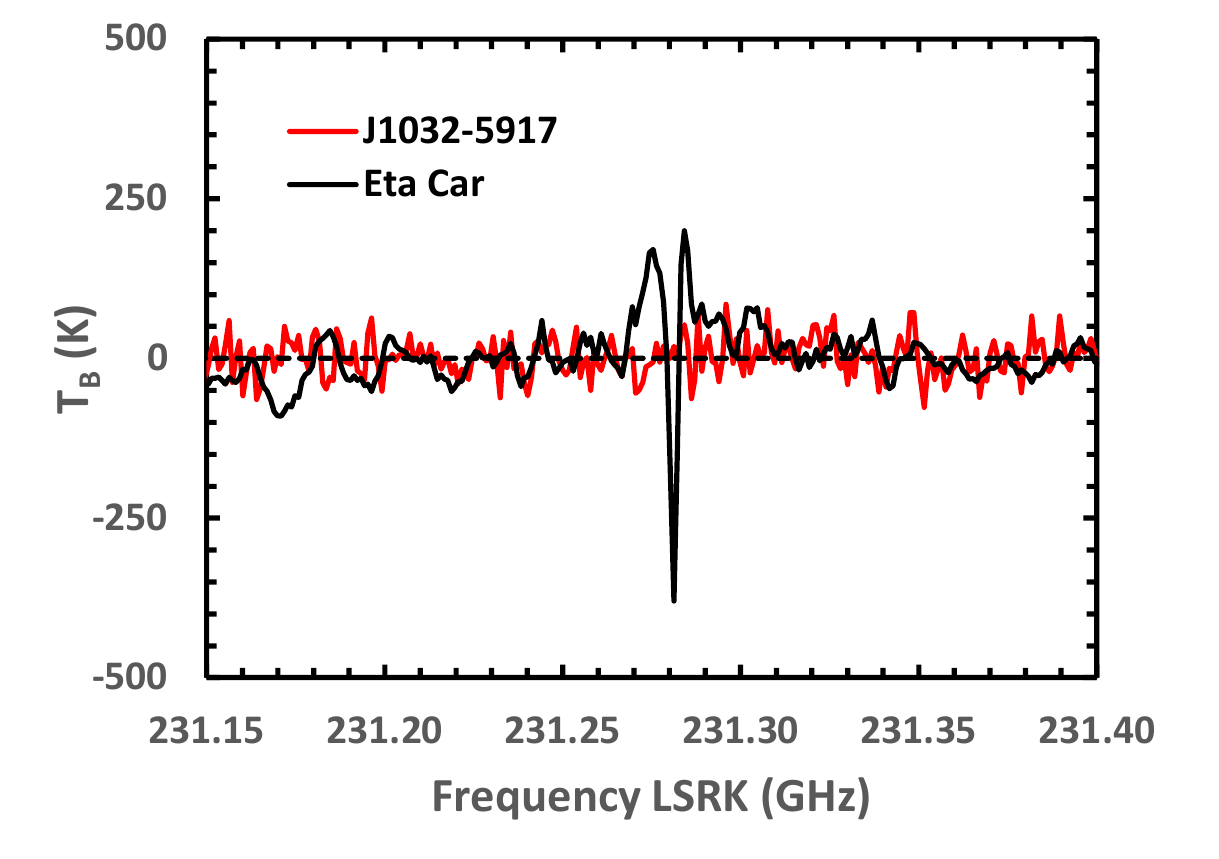}
\caption{Spectra showing the the absorption line at 231.28 GHz in $\eta$ Car obtained on 2017 November (black), and that of the corresponding phase calibrator source J1032-5917 (red). }
\label{fig:Fig_3_telluric.pdf}
\end{center}
\end{figure}

We re-analyzed our data of $\eta$ Car and of the phase calibrator at frequencies close to the frequency of the absorption line, to avoid the  H30$\alpha$ strong emission. We  used the CASA Hogbom algorithm to clean a data cube with the highest  possible velocity resolution, taking 350 channels around the 321.28 GHz  absorption line. We  used Briggs weighting with the robust parameter set to 0.5.
\begin{figure*}
\begin{center}
\includegraphics[width= 17cm]{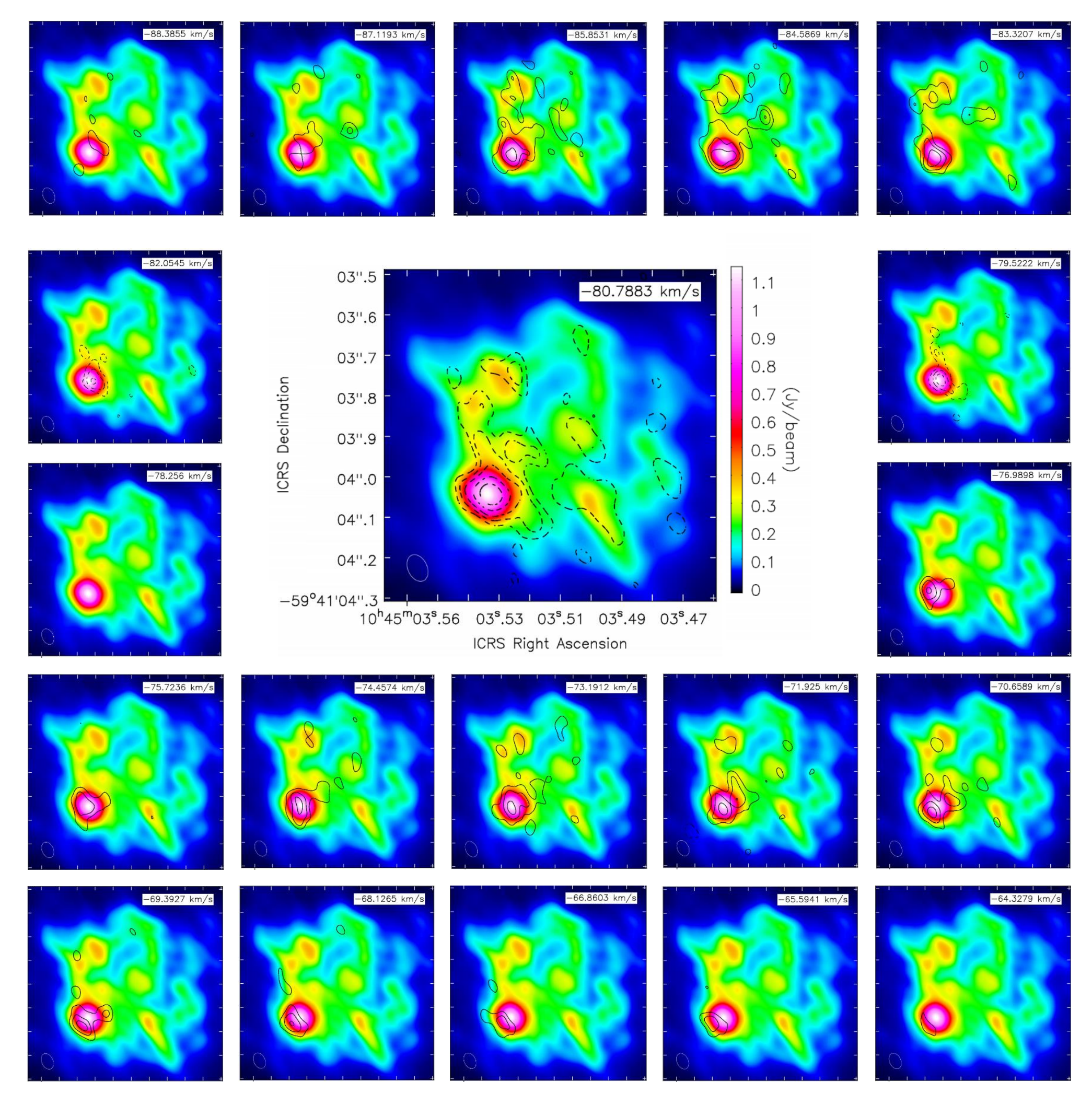}
\caption{Intensity maps of velocity channels, centered arbitrarily at the frequency  of the $^{13}$CS (5-4) line (zero velocity), with width of 1.262 km~s$^{-1}$,  showing   emission  (solid black contours) and absorption (broken black contours), superimposed to the 230 GHz continuum raster image. All the maps have the same intensity contours: -0.6, -0.4, -0.28, -0.18, 0.18, 0.28 and 0.4 of 0.0543 Jy beam$^{-1}$.
The central velocities of the  maps are indicated in the upper right corner of each panel. The beam size is represented by the white ellipse at the bottom left corner of each panel. The coordinate axis and continuum flux density scale are presented in the central large image. }
\label{fig:Fig_4_telluric.pdf}
\end{center}
\end{figure*}
\begin{figure*}
\begin{center}
\includegraphics[width=17cm]{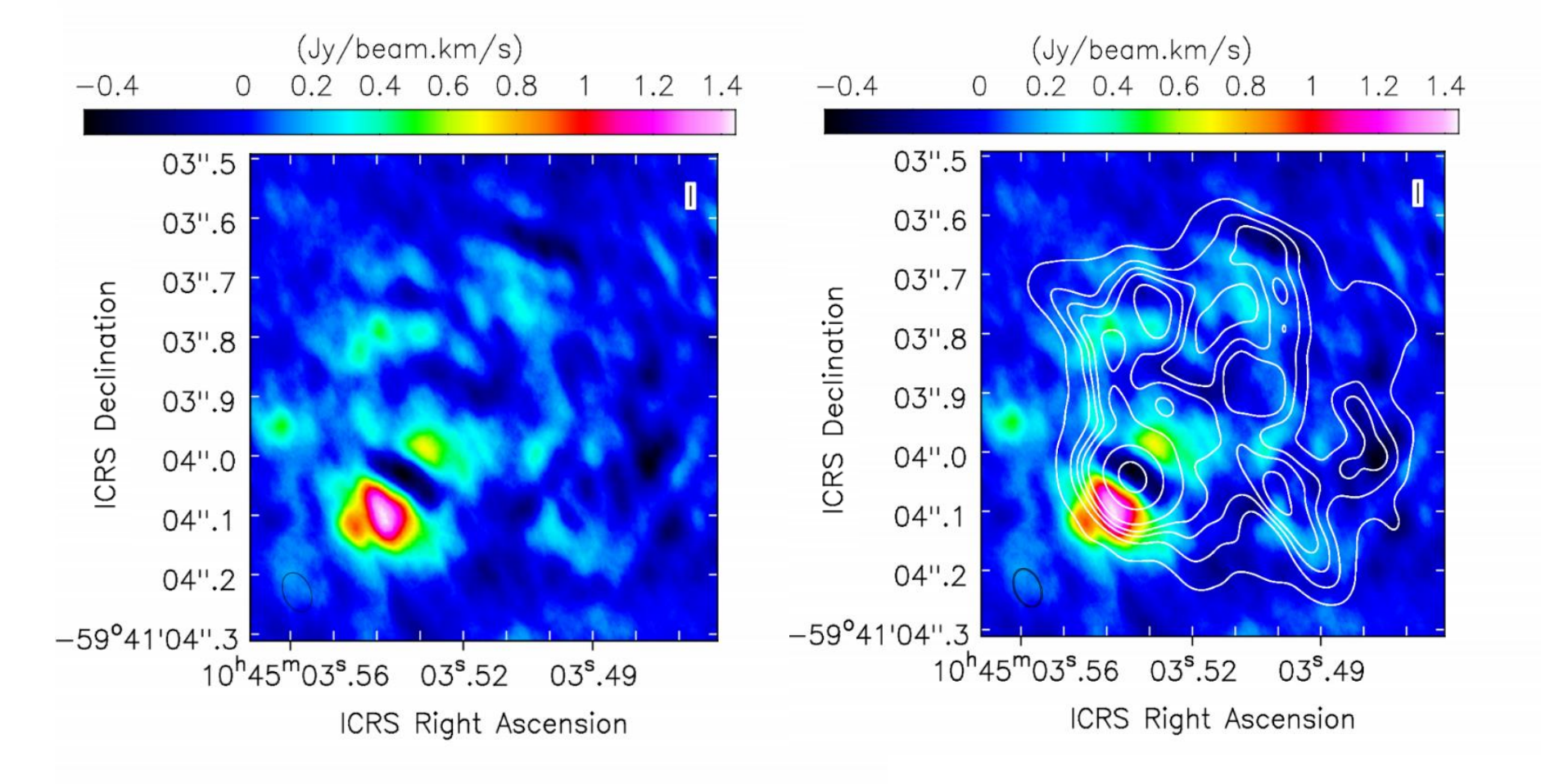}
\caption{Left panel: zero momentum raster image of the absorption line, centered arbitrarily at the frequency of the  $^{13}$CS (5-4) line (zero velocity), integrated  between $-260$ and 180 km s$^{-1}$. Right: same zero momentum raster image with the 230 GHz continuum intensity contours superimposed. The continuum intensity contours are 0.10, 0.17, 0.22, 0.36, 0.55 and 1.00 Jy beam$^{-1}$. The intensity scales of the raster images and the synthesized beams (as black contours) are displayed on the top  and  bottom left corner of each panel, respectively. }
\label{fig:Fig_5_telluric.pdf}
\end{center}
\end{figure*}

\begin{figure*}
\begin{center}
\includegraphics[width= 17cm]{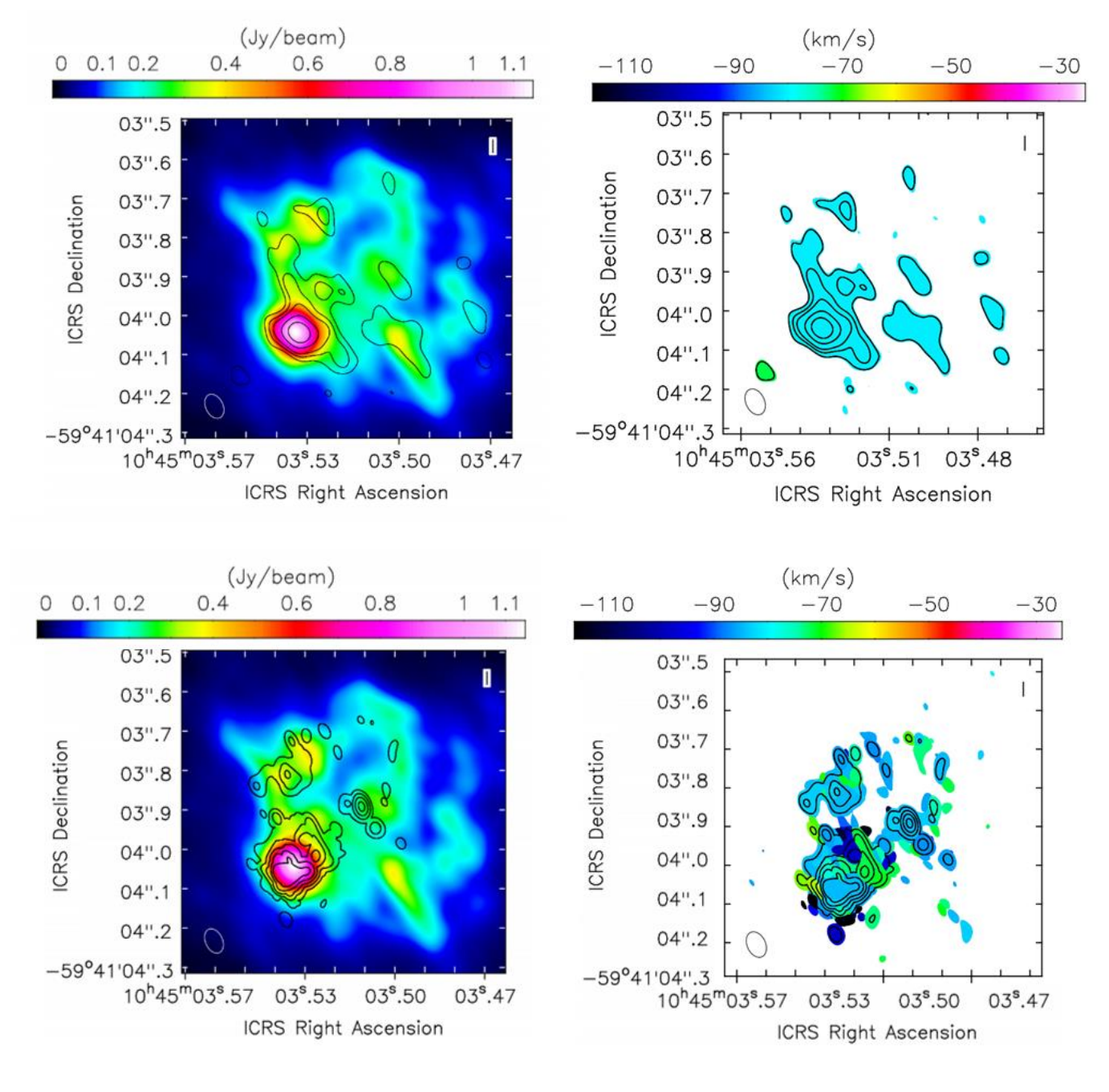}
\caption{Left panels (top/bottom): maximum intensity contour maps of the  of  231.28 GHz absorption/emission lines, superimposed to the raster image of the 230 GHz continuum. The contours of the intensity maps  are: -0.044, -0.033, -0.022, -0.016, and -0.011  Jy beam$^{-1}$ for absorption and  0.012, 0.015, 0.018, 0.021 and 0.024 Jy beam$^{-1}$ for emission.
The scales of the continuum raster images are shown at the top of each panel and the beam sizes are represented by the white ellipses at the bottom left of each panel. Right panels (top/bottom): the same contour maps as in the left panels, superimposed to the raster images of the velocities corresponding to the maxima in the absorption/emission intensity maps. The velocity scale, shown at the top of the panels,  is referred  arbitrarily to the  $^{13}$CS (5-4) transition zero velocity, the beam size is shown as a black contour at the bottom left of the panels.}
\label{fig:Fig_6_telluric.pdf}
\end{center}
\end{figure*}


We also re-analyzed the data in project 2013.1.00661.S extracted from the ALMA public archive, focused on the presence of the absorption lines centered at 231.28 GHz and close to the frequency of the CO~(2-1) transition.  The observations were obtained on 2015 April  with resolution ($1\farcs3 \times 0\farcs99$), using four spectral windows, all of them with the same spectral resolution (channel width of 488.281 kHz).  They were centered at 230.566, 232.521, 218.520, and 216.645 GHz, with a  total bandwidth of 1.875 GHz. Bandpass and phase calibrator sources were J1107-4449 and J1047-6217, respectively. As we did with the previous data-set, we  used the Hogbom algorithm of CASA to clean the data-cube. We obtained a cube with 1.26 km s$^{-1}$ velocity resolution, with a cell size of 50 mas and $(512\times 512)$  pixels around the phase centre. We set Briggs weighting with a robust parameter of 0.5.

Finally, we re-analysed the data of project 2016.1.00585.S, extracted also from the ALMA public archive. These observations were performed on 2016 October, using 4 different spectral windows, all of them with a bandwidth of 937.5 MHz, one with a channel width of 244 kHz, and the others with  channel widths of 488 kHz. The higher frequency resolution spectral window was centered at 345.816 GHz, while the others were centered at 345.361, 354.858, and 356.756 GHz. The bandpass and phase calibrator sources were J0538-4405 and J1047-6217, respectively.
The final data cubes were obtained using the Hogbom algorithm of CASA, once again using Briggs weighting with the robust parameter set to 0.5. We  used a pixel size of 20 mas in an image of $(512\times 512)$ pixels, which allows us to investigate the $(10'' \times 10'' )$  region centered on $\eta$ Car. 
\section{Results}
\label{sec:Results}

Figure \ref{fig:Fig_1_telluric.pdf} shows the continuum images of the three sets of observations discussed in this paper. The left panel image, at the frequency of 230 GHz, with resolution ($0\farcs065 \times 0\farcs043$), was obtained on 2017 November and  was already presented by \citet{abr20}; the central panel image, at the same frequency but with lower resolution ($1\farcs3 \times 0\farcs99$), corresponds to ALMA observations obtained in 2015 April,   as part of  CO~(2-1) observations \citepalias{smi18}. 
The right panel image, with resolution ($0\farcs17 \times 0\farcs12$) and frequency 354 GHz, obtained on 2016 October, was discussed by \citet{bor19}. 
These images are crucial for the understanding of the absorption features that are discussed in the remaining of this paper. 
The  flux densities at 230 GHz,  integrated over all the emission regions, are  $23.5\pm 0.1$ Jy and $28.30 \pm 0.01$ Jy on 2015 April and  2017 November, respectively, and $33.83 \pm 0.02$ Jy   at 354 GHz, on 2016 October. At the higher resolutions, it is possible to separate  the compact source, coincident with the position of $\eta$ Car, from the extended emission NW of the binary source, coincident with the Weigelt Complex.
\begin{figure*}
\begin{center}
\includegraphics[width= 17cm]{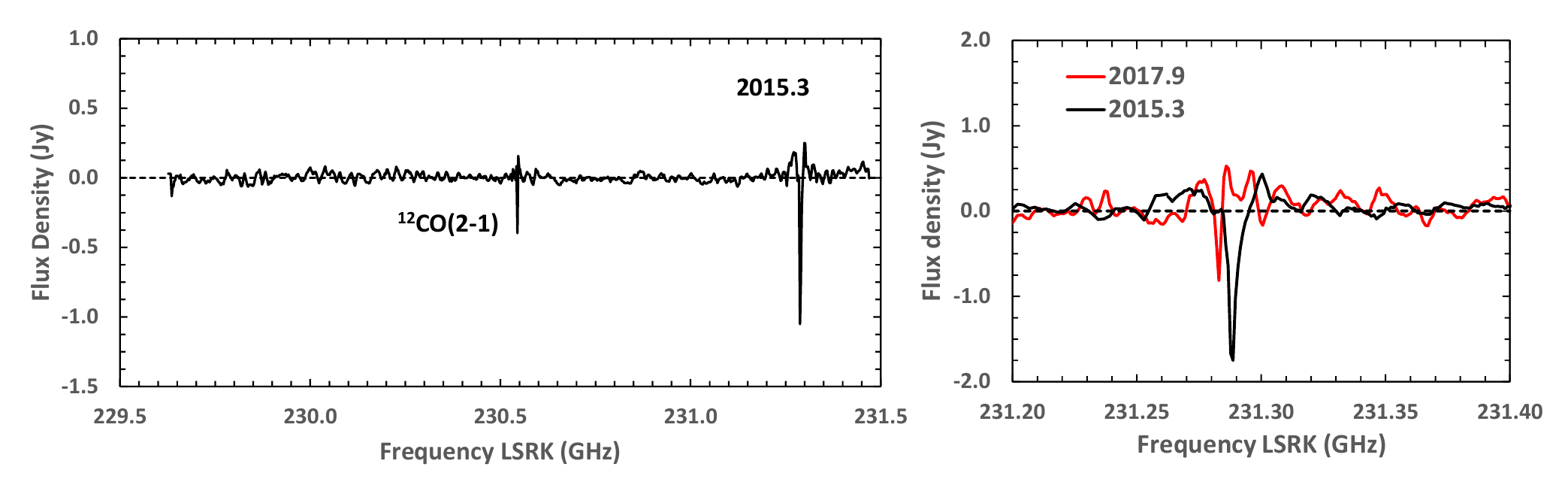}
\caption{Left: spectrum of the band centered at the rest frequency of the CO~(2-1) transition, with velocity resolution of 3 km s$^{-1}$, integrated over the region that covers the continuum emission obtained on 2015.3, which also shows the 231.28 GHz absorption line. Right: spectrum of the  231.28 GHz absorption line on two epochs: 2015.3 (red) and 2017.9 (black), with velocity resolution of 1.26 km s$^{-1}$, integrated over the regions that cover the continuum emission}  
\label{fig:Fig_7_telluric.pdf}
\end{center}
\end{figure*}
\begin{figure*}
\begin{center}
\includegraphics[width= 17cm]{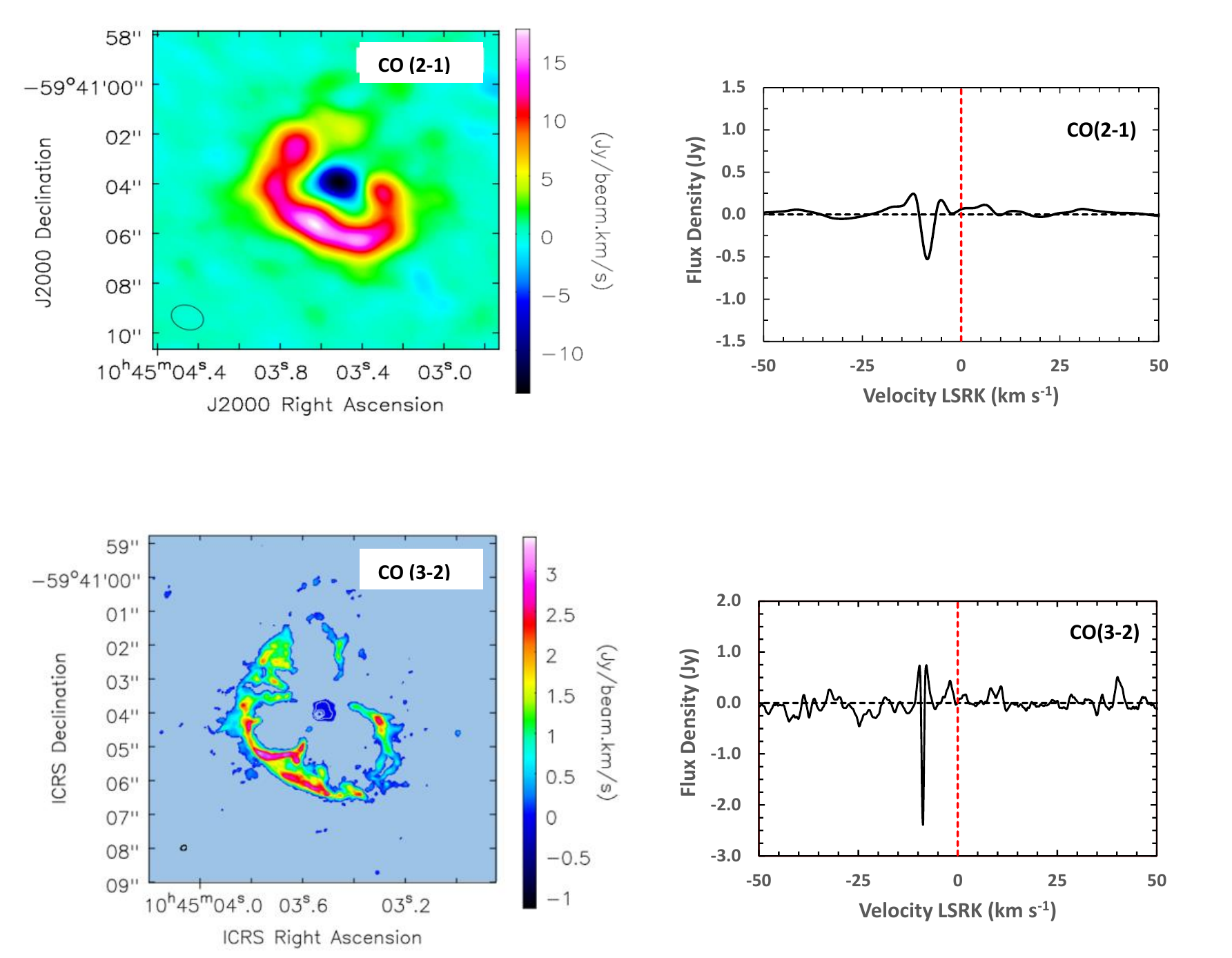}
\caption{Upper left panel: integrated intensity raster image of  the CO~(2-1) transition (zero momentum) obtained on 2015.3, with resolution indicated by the black ellipse at the bottom left of the image.  Top right panel:  spectrum of the CO~(2-1) transition, integrated over the continuum emission region.  Bottom left panel: integrated intensity raster image of  the CO~(3-2) transition (zero momentum) obtained on 2016.8, with resolution indicated by the black ellipse at the bottom left of the image. The white contours at the center of the image represent the continuum emission. Bottom right panel: spectrum of the CO~(3-2) transition, integrated over the continuum emission region}
\label{fig:Fig_8.pdf}
\end{center}
\end{figure*}
\begin{figure}
\begin{center}
\includegraphics[width= 8cm]{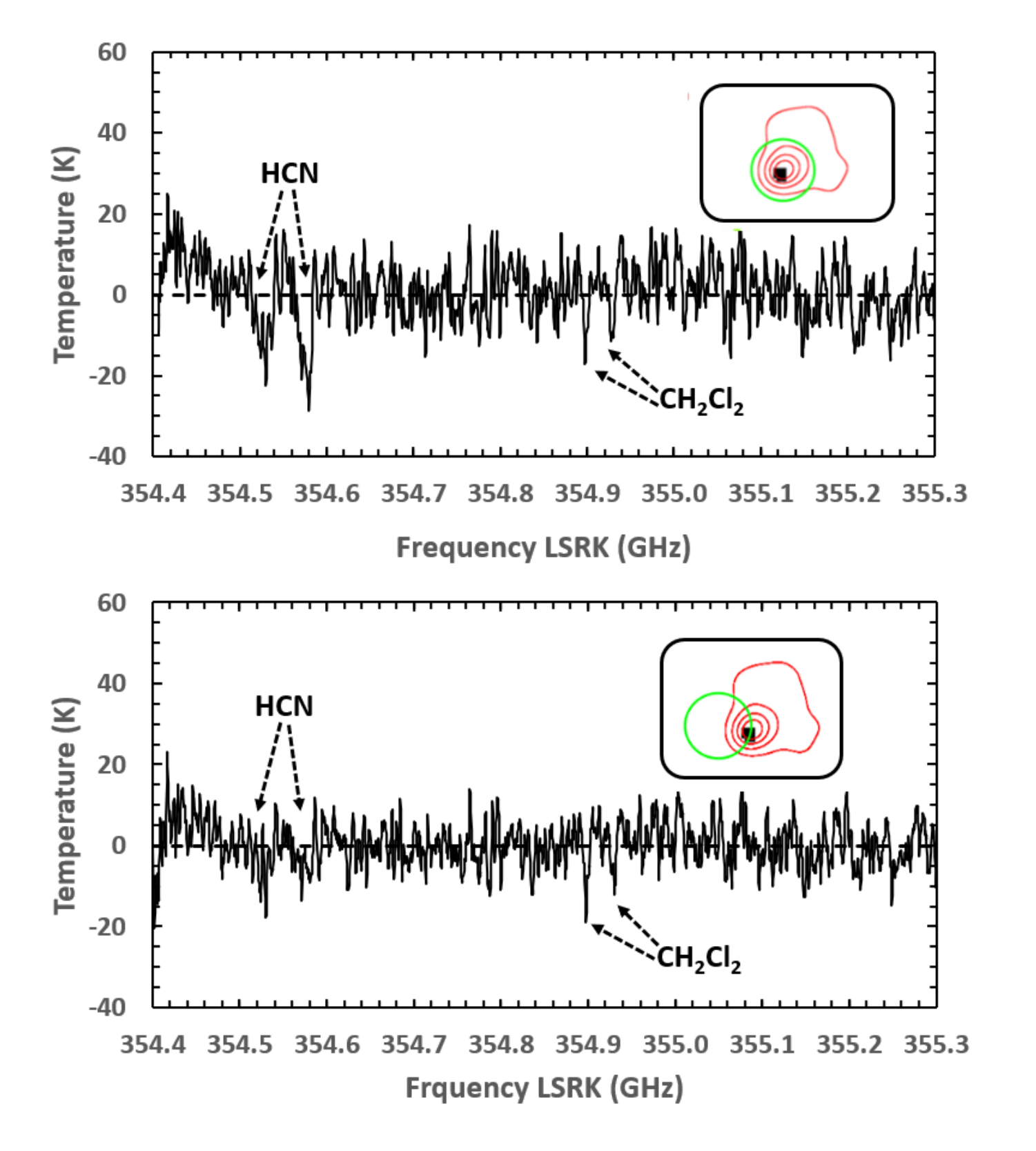}
\caption{Spectra of the band centered at the frequency 354.4 GHz obtained with ALMA on 2016.8, integrated over the regions delimited by the green circles in the top left insert of the respective panels, showing the HCN(4-3) lines in absorption and two other absorption lines at 354.90 and 354.93 GHz, which we identified as transitions of CH$_2$Cl$_2$. The red contours in the inserts represent the 354 GHz continuum emission.}
\label{fig:Fig_9.pdf}
\end{center}
\end{figure}
\begin{figure*}
\begin{center}
\includegraphics[width= 17cm]{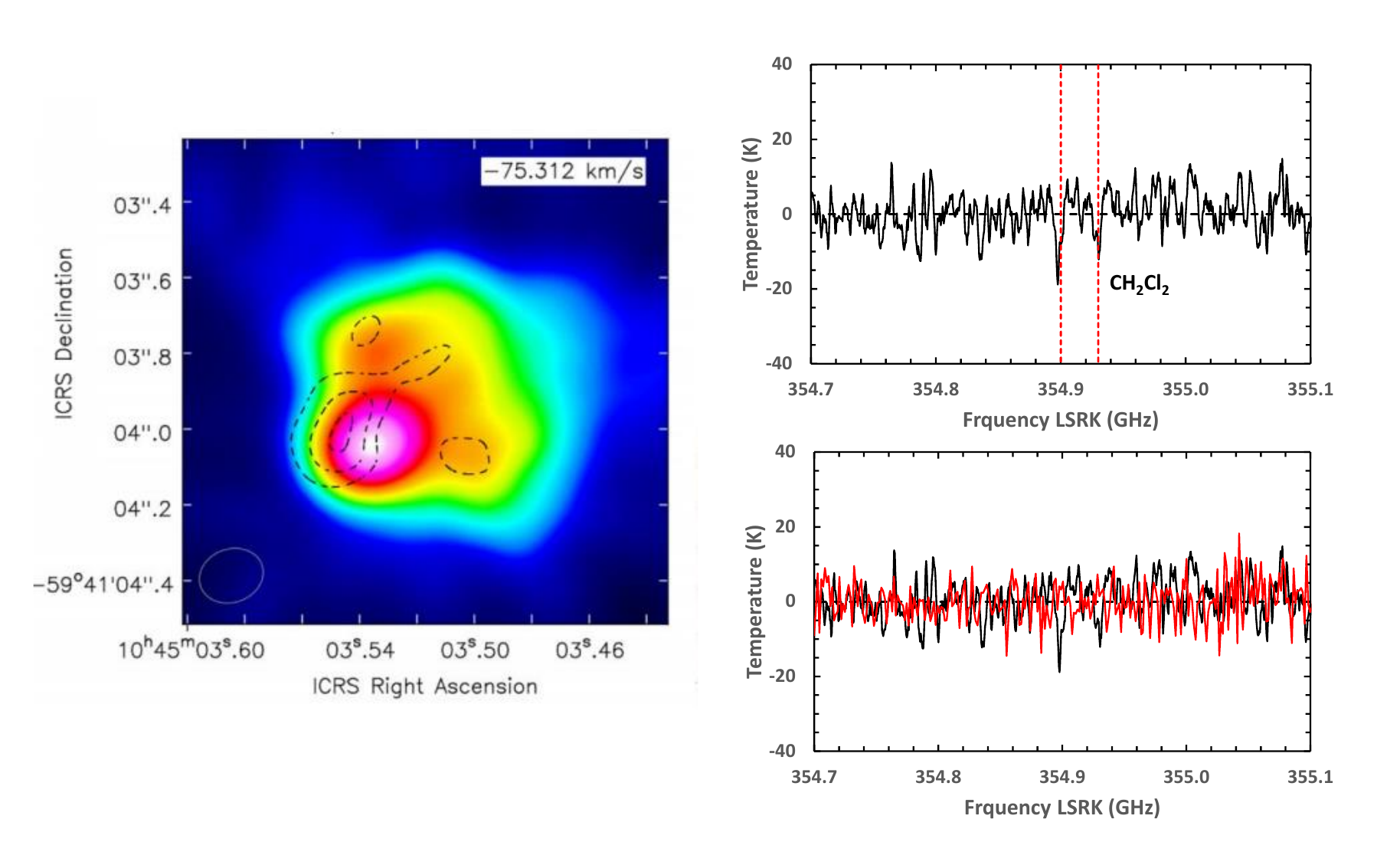}
\caption{Left panel: Intensity map of the velocity channel  at which the absorption of the 354.86 GHz line is maximum,   obtained on 2016.8 with resolution of 1.3 km~s$^{-1}$ (broken black contours), superimposed to the 354 GHz continuum raster image. The intensity contours are: -0.07, -0.05, -0.03, 0.03, 0.05, and 0.07 Jy/beam.
 The LSR velocity of the line is indicated in the upper right corner of the panel; zero  velocity corresponds to the frequency of 354.80 GHz. Top right panel: spectrum of the 354.86 and 354.92  lines in $\eta$~Car integrated over the whole continuum image; the vertical lines correspond to the frequencies of two transitions of the CH$_2$Cl$_2$ molecule. Bottom right panel: spectra of $\eta$~Car (black) and of the phase calibrator J1032-5917 (red), integrated over the whole continuum image. }
\label{fig:Fig_10_telluric.pdf}
\end{center}
\end{figure*}
\begin{figure}
\begin{center}
\includegraphics[width= 8cm]{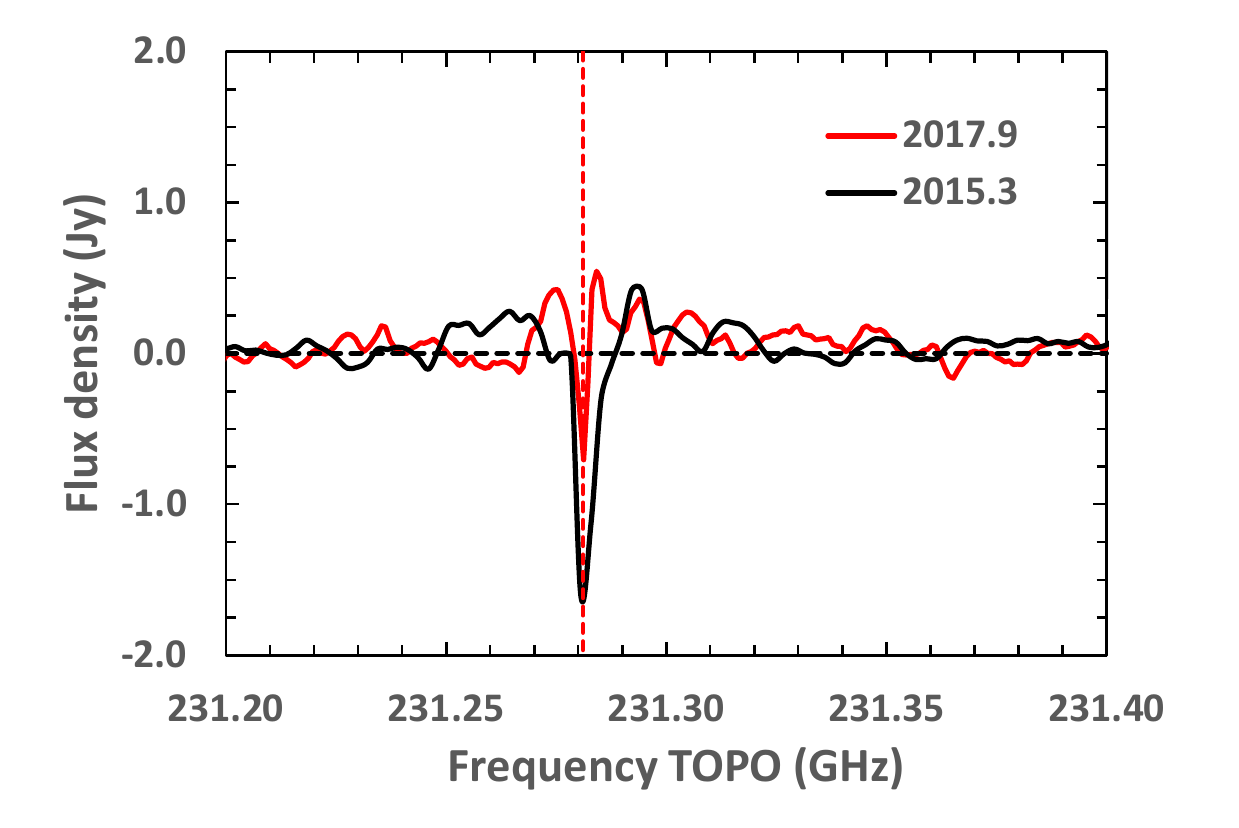}
\caption{The same spectra as presented in the right panel of Fig. \ref{fig:Fig_7_telluric.pdf} but with the frequency axis in the topocentric reference frame. }
\label{fig:Fig_11.pdf}
\end{center}
\end{figure}
\begin{figure*}
\begin{center}
\includegraphics[width= 17cm]{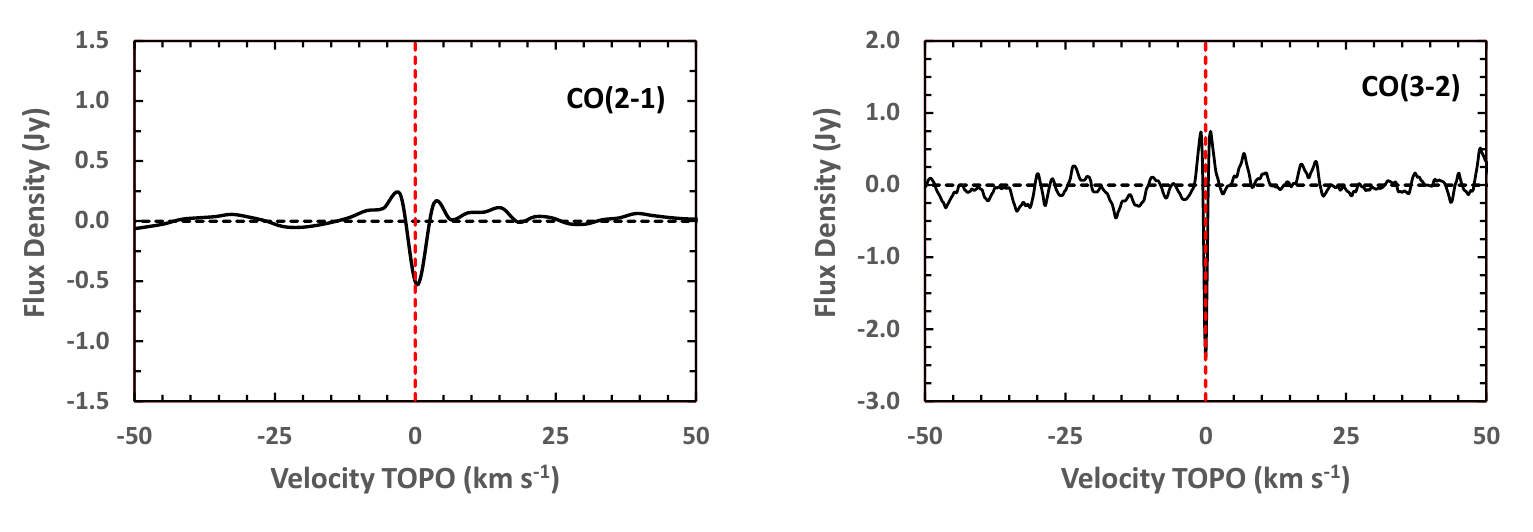}
\caption{The same CO~$(2-1)$ and CO~$(3-2)$ spectra of Fig. \ref{fig:Fig_8.pdf} but with the frequency axis in the topocentric reference frame.}
\label{fig:Fig_12.pdf}
\end{center}
\end{figure*}
\subsection{Absorption line at 231.28 GHz }
\label{CShigh}

Figure \ref{fig:Fig_2_telluric.pdf} (left panel) presents the  spectrum of $\eta$ Car with  1.26 km s$^{-1}$ velocity resolution, obtained with ALMA on 2017 November, with ($0\farcs065\times 0\farcs043$) spatial resolution , integrated over the whole continuum emitting region shown in the left panel of Fig. \ref{fig:Fig_1_telluric.pdf}. It shows the strong H30$\alpha$ and He30$\alpha$ recombination lines in emission and a weak absorption line at the frequency of 231.28 GHz,  displayed on an enlarged flux density scale in the insert at the top  left corner of the panel. The recombination lines were already discussed by \citet{abr20}: the H30$\alpha$ line was resolved in at least 16 compact components, with velocities ranging from  $-30$ to $-70$ km s$^{-1}$; they are located in a $0\farcs6$ region NW of $\eta$ Car, known as the Weigelt Complex. Those are bright emission lines with no coincidence with any telluric line emission.

The right panel of Figure \ref{fig:Fig_2_telluric.pdf}, shows the expanded spectrum of the 231.28 GHz absorption line integrated over the continuum emitting region. \citet{mor20} identified this absorption line, also present in   observations of the  CO disrupted torus surrounding $\eta$ Car \citepalias{smi18}, as the methanol transition H$_3$COH~($10_2-9_3)A^-$,  with  LSR velocity close to zero, although the $^{13}$CS~(5-4) transition was also mentioned as a possible identification, in which case, the velocity of the absorbing cloud would be about -80 km s$^{-1}$, while the velocity of $\eta$ Car in this reference frame is -19.7 km s$^{-1}$. The vertical lines in the figure show the rest frequencies of the $^{13}$CS~(5-4), H$_3$COH~($10_2-9_3)A^-$ and telluric O$_3$~$[16(1,15)-16(0,16)]$ transitions.

To check if the absorption line is the result of a faulty calibration, we analysed the phase calibrator J1032-5917, which is close to $\eta$ Car  in the sky and obtained its continuum and spectral images. J1032-5917 and $\eta$ Car were observed almost simultaneously  and used the same bandpass calibrator. As expected, the continuum image of J1032-5917 is not resolved; its  flux density is $0.22 \pm 0.01$ Jy. The spectrum, together with that of $\eta$ Car is shown in Figure \ref{fig:Fig_3_telluric.pdf};  both spectra were integrated over the same continuum emitting region  shown in the left panel of Fig. \ref{fig:Fig_1_telluric.pdf}; the vertical axis is now the brightness temperature  $T_{\rm b}(\nu)$, defined as:
 
\begin{equation}
    T_{\rm b}(\nu)=1.22\times 10^6\frac{S_{\nu}}{{\nu}^2 \theta_{\rm maj} \theta_{\rm min}}{\rm   K},
\end{equation}

\noindent
where $S_{\nu}$ is the flux density in Jy beam$^{-1}$, $\nu$ the frequency in GHz,  $\theta_{\rm maj}$ and $\theta_{\rm min}$ the major and minor axis of the elliptical beam in arcsec. We can see that although the noise levels in both spectra are similar, the absorption line is not present in the calibrator source.

To further analyse the origin of the absorption line, in Figure \ref{fig:Fig_4_telluric.pdf}  we present the intensity contour maps of each velocity channel in which the line is present in absorption or emission, superimposed to the 230 GHz continuum raster image.  The image cube was centered arbitrarily  at the frequency of the $^{13}$CS~(5-4) line (zero velocity). Although  we do not know it the line is intrinsic to the source, differences in velocity are used instead of differences in  frequency because they are  easier to understand.



Figure \ref{fig:Fig_5_telluric.pdf} presents  intensity raster  images of  the absorption line,  integrated  between the velocities of $-260$ and $180$ km s$^{-1}$ (zero momentum). A noise level corresponding to $\pm 3$ mJy beam$^{-1}$ was excluded from the integration.
The image in the left panel shows a narrow absorption ridge, in the NE-SW direction; the image on the right panel  also includes the contour map of the 230 GHz continuum emission, showing  the position of the ridge relative to the continuum source.

Since line emission and absorption may cancel out in the integrated intensity image, we obtained also images of minimum and maximum intensity representing absorption and emission respectively, and also images of the velocities at which  these extreme intensities occur.
Figure \ref{fig:Fig_6_telluric.pdf} left panel (top/bottom) presents the contour maps of the maximum intensity of the  line  in absorption/emission, superimposed to the raster image of the  230 GHz continuum.  
The right panels (top/bottom) of Figure \ref{fig:Fig_6_telluric.pdf} present the contour maps of the maximum intensity of the line in absorption/emission, superimposed to the raster images of the velocities corresponding to these maxima.


The agreement between the absorption contours at the centre of the absorption line ($-80.8$ km s$^{-1}$) and the continuum image  showed in figure \ref{fig:Fig_4_telluric.pdf} is remarkable,  covering both  $\eta$ Car and  the Weigelt Complex region. The 231.28 GHz emission line intensity contours are  only seen in front of the continuum source, implying that, if the lines are real, they originate in a cloud that only covers the continuum source.


The absorption line at 231.28 GHz  was also found in  observations obtained with ALMA on 2015 April (2015.3), with spatial resolution ($1\farcs33 \times 0\farcs99$). These observations were retrieved from the ALMA public archive and processed using a CLEAN window of $13\arcsec$ and velocity resolution of 1.26 km s$^{-1}$.

Figure \ref{fig:Fig_7_telluric.pdf} (left panel) presents the spectrum of the band centered at the frequency of the CO~(2-1) line, with velocity resolution of 3 km s$^{-1}$, integrated over the region that covers the continuum source (central panel of Fig. \ref{fig:Fig_1_telluric.pdf}). The CO~(2-1) and the 231.28 GHz lines can be seen in absorption at the centre and high frequency end of the band, respectively. The right panel of Figure  \ref{fig:Fig_7_telluric.pdf} presents together the 231.28 GHz lines, observed on 2015.3 and on 2017.9, with 1.26 km s$^{-1}$ velocity resolution. We can see that the line was wider on 2015.3 than on 2017.9, and that the central frequency of the lines differ in an amount that corresponds to approximately 8 km s$^{-1}$.
\subsection{Absorption line close to CO~(2-1) and CO~(3-2) lines}
\label{CO line}
In this section we compare the absorption lines observed in front of the 230 and 354 GHz continuum (centre and right panel of Fig. \ref{fig:Fig_1_telluric.pdf}), close to the frequency of the CO~(2-1) and CO~(3-2) transitions, respectively.  

The top left panel of Figure \ref{fig:Fig_8.pdf}  shows the raster image of the CO~(2-1) transition, integrated over the whole spectrum (zero momentum);  we can see the CO torus in emission and strong absorption in front of the central continuum source. The top right panel of Fig. \ref{fig:Fig_8.pdf} shows the absorption line, with a velocity resolution of 3 km s$^{-1}$; the LSR velocity at the center of the line is  -8.54 km s$^{-1}$.

The CO~(3-2) absorption line was reported by \citet{bor19} as part of the study of the disrupted $2 \arcsec$ radius torus seen in emission. We re-analyzed the data and show the image of the CO~(3-2) line, integrated over the whole spectrum (zero momentum) at the bottom left panel of Figure \ref{fig:Fig_8.pdf}, and the absorption line, with velocity resolution of 1.3 km s$^{-1}$ at the bottom right panel of the same figure. The LSR velocity at the center of the line is -8.74 km s$^{-1}$, coincident with the velocity of the CO~(2-1) transition observed about one year earlier.

\subsection{The 354.90 and 354.93 GHz absorption line}
\label{arcs}

Besides the lines mentioned above, we also detected two narrow absorption lines  in the band centered at 354.8 GHz, in data retrieved from ALMA public archive  obtained on 2016 October. These data were used by \citet{bor19} to study the HCN~(4-3) transition, seen in emission in the $2\arcsec$ radius torus that surrounds the $\eta$ Car binary system and also in absorption against the central continuum source, at the frequencies of 354.51 and 354.46 GHz.
The absorption lines reported here are weak but very well defined; they are  centered at the frequencies of 354.90 and 354.93 GHz with a line width of about 1.3 km s$^{-1}$. These frequencies are close to two telluric lines of the of the methyl chloride CH$_2$Cl$_2$ molecule.

Spectra of the whole band are presented in Figure \ref{fig:Fig_9.pdf}. They include the HCN~(4-3) lines reported by \citet{bor19} and the newly reported absorption lines.
The top panel of Fig. \ref{fig:Fig_9.pdf} shows the spectrum integrated over the continuum emission region  delimited by  the green circle superimposed on the red 354 GHz continuum contours, on the insert at the top left of the panel.  In this region, the HCN lines are stronger than the absorption lines reported here. 
The bottom panel shows the spectrum, integrated over the same area than that in the top panel, but in a position centered to the east of the compact continuum source, as shown by the green circle at the insert in the top left of the  panel. We can see that the  intensity of the absorption lines reported here remain the same as in the previous region, while the HCN line intensity had decreased, implying that if the reported lines are real, they probably  originate in a more extended region than HCN.

Figure \ref{fig:Fig_10_telluric.pdf} (left panel) shows the intensity contour map of the velocity channel in which 354.89 GHz  line  presents maximum absorption, superimposed on the 354 GHz continuum raster image. We can see that the absorption is stronger towards the the east of the compact continuum source, as already shown in the integrated spectra of Fig. \ref{fig:Fig_9.pdf}. Since the 354.93 line is weaker it is not shown here; it is centered on the continuum source but it does not present any structure.

Once again, we analysed  the spectrum of the phase calibrator source, observed almost simultaneously with $\eta$~Car. At the top right panel of Figure \ref{fig:Fig_10_telluric.pdf} we display the spectrum of the absorption lines detected in  $\eta$~Car, showing the frequencies of the two transitions of the CH$_2$Cl$_2$ molecule and at the bottom right panel we show the spectrum of  $\eta$~Car together with the spectrum of phase the calibrator source J1032-5917. Both sources used the same bandpass calibrator J0904-5735. Although the noise level in the phase calibrator and in $\eta$~Car are similar, the spectrum of the calibrator does not present any absorption feature that can be related to a telluric molecule.
\begin{figure*}
\begin{center}
\includegraphics[width= 17cm]{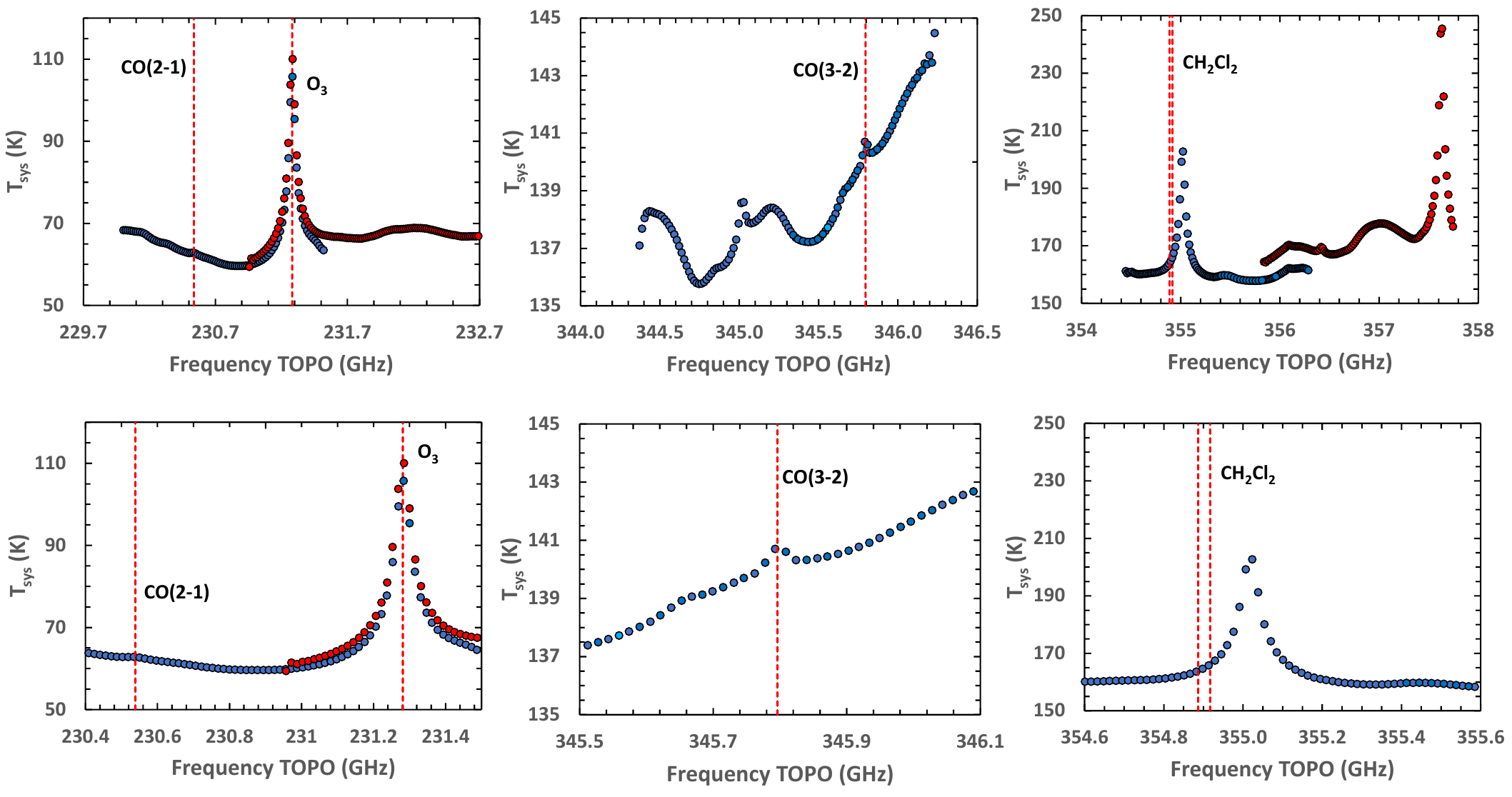}
\caption{Top: System temperature of the different sets of  observations, averaged over all the antennas, as a function of frequency. Bottom: the same Tsys, in an expanded frequency scale. Vertical lines represent the position of the telluric lines that appeared as absorption lines in the target sapectra. }
\label{fig:Fig_13.pdf}
\end{center}
\end{figure*}

\section{Discussion}
\label{Discussion}

In Section \ref{sec:Results} we presented images of the central region of $\eta$~Car in the 230 and 354 GHz continuum with different spatial resolutions: ($0\farcs065 \times 0\farcs043$), ($1\farcs3 \times 0\farcs99$) and ($0\farcs17 \times 0\farcs12$), obtained on three different epochs, 2015.3, 2016.8 and 2017.9, respectively. We also showed their spectra, focusing on the presence of absorption lines with LSR frequencies close to the frequencies of  telluric lines of the molecules  O$_3$, CO and CH$_2$Cl$_2$.

To understand the possible contamination of the target spectra by telluric lines, we must consider the calibration procedure used by ALMA. The observations of the target and phase calibrators are made in frequency division multiplexing mode (FDM), which allows frequency resolution of up to 3840 channels, but the bandpass calibrator is observed in  time multiplexing mode (TDM), which provides only 128 frequency channels. Therefore, the bandpass correction does not take into account the contribution of all the individual channels, but only the mean value provided by the bandpass calibrator.

\citet{hun18} investigated the effect of this calibration procedure when  telluric lines are present, in particular the very strong ozone lines, and showed that they could contaminate the spectrum of a line-free  source, specially when the   the bandpass calibrator and the target source are observed at different elevations. To verify if this was the case in the observations discussed in this paper, we present in Table \ref{tab:Table_1} the conditions at which the observations of the target, bandpass and  phase calibrators  were made (name of the bandpass and phase calibrators and elevation of the sources at the date, UTC and ST times of observations. We can see that the differences in elevation between the bandpass calibrators and the target were $14^\circ$, $22^\circ$ and only $2^\circ$ on 2015.3, 2016.8 and 2017.9, respectively. Therefore, at least on the last epoch we would not expect to find the contamination related by \citep{hun18}.

 The definitive proof that the lines are from telluric origin rests on the evaluation of their topocetric frequencies at the time of the observations, which are easily obtained from the CASA software. These frequencies are presented in Table \ref{tab:Table_2}, where we show the LSRK and the corresponding topocentric (TOPO) frequencies at the centre of the absorption lines, the velocity corresponding to the difference between the two frequencies, which represents the earth velocity, as well as the frequencies of the corresponding tellutic lines and their difference with the topocentric frequency of the absorption lines.
 
 In the case of the 231.28 GHz line, the topocentric frequencies at the centre of the absorption lines  at the two epochs coincide with the frequency of the O$_3$~$[16(1,15)-16(0,16)]$ transition, while their LSRK frequencies were different. This can be also seen in Figure \ref{fig:Fig_11.pdf} that presents together the two lines, observed on 2015.3 and 2017.9, in the topocentric reference frame, showing that the difference in the LSRK frequency at the center of the lines shown in Fig. \ref{fig:Fig_7_telluric.pdf} was due to differences in the velocity of the earth relative to the LSRK,  confirming their telluric origin.
 
The CO~(2-1) and CO~(3-2) transitions,  were observed on different epochs but presented similar velocities relative to the rest frequencies of the CO molecule in the LSRK reference system; their topocentric frequencies correspond exactly to the rest frequency of the respective transitions, which can be considered also a strong evidence of their telluric origin. Figure \ref{fig:Fig_12.pdf} presents the spectra of the two transitions in the topocentric velocity reference frame, showing that both lines are centered at zero velocity.

The two low intensity absorption lines observed at frequencies of 354.90 and 354.93 GHz were observed only on one epoch, but their topocentric frequencies coincide with the frequencies of two CH$_2$Cl$_2$~$[6(6,0)-5(5,1)]$~$I=1-3,F=6-7$ and  CH$_2$Cl$_2$~$[6(6,0)-5(5,1)]$~$I=1-3,F=7-8$ transitions, which can also be considered a strong evidence of their telluric origin.

The frequencie of the two HCN absorption lines discussed by \citet{bor19} are not close to any strong telluric line and their profiles are very different from those of the lines we are discussing in this paper, but the strongest reason to believe that they are real is that the frequency difference between them corresponds to the difference between two HCN transitios HCN~(J=4-3, F = 3-4) and HCN~(J=4-3, F=4-4, I=1e).

\begin{table}
    \centering
    \begin{tabular}{l|c|c|c}
    \hline
    Object &   Bandpass  &   Phase    & Target \\
     &  Calibrator & Calibrator &    \\
     \hline
     \hline
    Date  & &  3/04/2015 & \\
     \hline
     Source & J1107-4449 & J1047-6217 & $\eta$~Car \\
     RA  (ICRS)~~  $\rm {(h:m:s)}$ & 11:07:08.69  & 10:47:42.94   & 10:45:03.54  \\
     DEC (ICRS) ($^\circ:~\arcmin~:\arcsec$) & -44:49:07.62 & -62: 17:14.65 &  -59:41:04.05 \\
     UTC~$\rm {(h:m)}$ & 3:38 & 3:48  &  3:48 \\
     ST~~~ $\rm {(h:m)}$ & 10:46 & 10:57 & 10:58 \\
     El $(^\circ)$ & 67.7 & 50.7 & 53.2 \\
     \hline
      \hline
    Date &  &  24/10/2016 & \\
     \hline
     Source & J0538-4405 & J1047-6217 & $\eta$~Car \\
     RA  (ICRS)~~  $\rm {(h:m:s)}$ & 05:38:50.36  & 10:47:42.94   & 10:45:03.54  \\
     DEC (ICRS) ($^\circ:~\arcmin~:\arcsec$) & -44:05:08.94 & -62:17:14.64 &  -59:41:4.05 \\
     UTC~$\rm {(h:m)}$ & 10:29 & 10:49  &  10:55 \\
     ST~~~ $\rm {(h:m)}$ & 07:07 & 07:26 & 07:32 \\
     El $(^\circ)$ & 62.6 & 38.2 & 40.3 \\
     \hline
    \hline
    Date &  &  20/11/2017 & \\
     \hline
     Source & J0904-5735 & J1032-5917 & $\eta$~Car \\
     RA  (ICRS)~~  $\rm {(h:m:s)}$ & 09:04:53.18  & 10:32:42.53   & 10:45:03.54  \\
     DEC (ICRS) ($^\circ:~\arcmin~:\arcsec$) & -57:35:05.78 & -59:17:57.12 &  -59:41:04.05 \\
     UTC~$\rm {(h:m)}$ & 11:21 & 11:35  &  11:26 \\
     ST~~~ $\rm {(h:m)}$ & 9:44 & 10:00 & 09:51 \\
     El $(^\circ)$ & 54.6 & 53.1 & 52.2 \\
     \hline
   
     \end{tabular}
     \caption{Position of $\eta$~Car, bandpass and phase calibrators during the observations}
    \label{tab:Table_1}
\end{table}

The last question that has to be answered is why, if the absorption lines are of telluric origin, they  are not detected in the phase calibrator sources, as shown in Figs. \ref{fig:Fig_3_telluric.pdf} and \ref{fig:Fig_10_telluric.pdf} of Section \ref{sec:Results}, which show the spectra of $\eta$~Car and of the phase calibrator sources (quasars), which were observed almost simultaneously with the target source and with the same spectral resolution (FDM),  but did not show any absorption line. The answer is that $\eta$~Car is a much stronger continuum source than the calibrators: 28 Jy at 230 GHz on 2017.9 compared with 0.2 Jy of the calibrator, and 33 Jy at 354 GHz on 2016.4, compared with 0.5 Jy of the calibrator.  That also explains the coincidence between the absorption lines and the continuum emission maps. It should also be noticed that the target source, $\eta$~Car, is  extended while the quasars are  point  sources, so that the comparison between them should be done at the level of flux density per beam. In  $\eta$~Car, on 2017.9, the maximum intensity of the continuum image at 230 GHz was 1.1 Jy beam$^{-1}$, while the maximum intensity at the centre of the absorption line was 0.054  Jy beam$^{-1}$, that is about 5\% of the continuum value. If the same ratio is applied to the 0.2 Jy beam$^{-1}$ of the calibrator source,  we obtain 0.01 Jy beam$^{-1}$ for the absorption line, which coincides with the maximum observed noise. Moreover, when   the spectrum of $\eta$~Car is integrated over the whole continuum emitting region, the systematic  errors in the calibration procedure add up, resulting in 0.78 Jy for the total flux density of the absorption line, while the integration of the phase calibrator point source  spectra  remains the same.

\begin{table}
    \centering
    \begin{tabular}{l|c|c|c|c|c}
    \hline
      Line   & Date Obs. & $\nu$ (LSRK) & $\nu$ (TOPO) & $\Delta V^{\rm a}$ & $\Delta \nu^{\rm b}$ \\
        &   &   GHz & GHz & km s$^{-1}$  & MHz \\
     \hline
      231.28 GHz & 2015.3 & 231.2875 & 231.2806 & 8.98 & 0.9 \\
      O$_3$ & & & 231.2815\\
     & \\
     231.28 GHz & 2017.9 & 231.2830 & 231.2812 & 2.30 & 0.3 \\
     O$_3$ & & & 231.2815\\
     \hline
     230.54 GHz & 2015.3 & 230.5446 & 230.5382 & 8.58 & -0.2 \\
     CO~(2-1) & & & 230.5380 \\
     & \\
     345.80 GHz & 2016.8 & 345.8061 & 345.7962 & 8.54 & -0.2 \\
     CO~(3-2) & & &  345.7960 \\
    \hline
    354.90 GHz & 2016.8 & 354.8977 & 354.8868 & 8.84 & -0.2 \\
    CH$_2$Cl$_2$ & & & 354.8870 \\
    & \\
    354.93 GHz & 2016.8 & 354.9275 & 354.9171 & 8.84 & 0.0 \\
    CH$_2$Cl$_2$ & & & 354.9171 \\
    \hline
    \end{tabular}
     \caption{Frequencies of the absorption and telluric lines in the LSRK and TOPO reference systems. (a) Velocity difference between LSRK and TOPO systems for $\eta$~Car; (b) difference between topocentric  frequency of $\eta$~Car and the frequency of the corresponding telluric line }
    \label{tab:Table_2}
\end{table}

Finally, it is worth discussing the best way to identify if an absorption line detected in an ALMA cube is real or an artificial feature due to a faulty subtraction of the telluric lines. 
A step that certainly helps the analysis is  looking carefully at the system temperature (T$_{\rm sys}$). In Figure \ref{fig:Fig_13.pdf} we show the system temperature for the data discussed in this paper.  To make those plots, we averaged the data through all antennas and during the entire observation. Since $\eta$~Car  is a bright source, the observation did not take too long, and there were no significant  changes in elevation.  The dashed red lines indicate the topocentric frequencies of the absorption lines discussed in this paper. We can see that T$_{\rm sys}$ at the frequency of the the O$_3$ lines is very large, but in others cases it does not have a pronounced change with frequency, as at the frequency of the CO(2-1) or CH$_2$Cl$_2$ lines. Therefore, the only direct way to verify if a line is telluric is to observe it at two epochs, in which the contribution of the earth velocity is different, and see if the LSR velocity of the line changes, while the topocentric velocity remains constant, as it was the case of the ozone line in $\eta$ Car. The availability of a list of the telluric lines modeled by the atmospheric calibration program would also help.

\section{Conclusions}
\label{sec:Conclusions}

Images of the region responsible for the 231.28 GHz absorption line observed in front of the strong continuum source in $\eta$~Car, obtained by ALMA with high spatial resolution on 2017 November, show strong similarity with the continuum image, which includes  the central source  and the Weigelt Complex.

The fact that the frequency of the absorption line is very close to the frequency of of a strong telluric ozone line and the report of \citet{hun18} about a possible inaccuracy in the calibration procedure in the presence of strong telluric lines, motivated a  more profound study of the origin of the observed line, as well as of other  lines  in absorption detected in data retrieved from ALMA public archive. These lines were: 231.28 GHz and CO~(2-1) observed on 2015 April, CO~(3-2), 354.89 and 354.92 GHz observed on 2016 October.

Since the inaccuracy in the calibration affected in the same way the phase calibrator and the target ($\eta$~Car) sources, we also obtained images and spectra of the phase calibrators; although the noise levels of their spectra were similar to those of the target source, no absorption lines were detected.

The  definite proof of the telluric origin of the absorption lines came from the calculation of their topocentric frequencies,  since their difference with the LSR frequencies reflect the velocity of the earth  relative to the LSR, which depends on the epoch of the observation. In all cases the topocentric frequencies  coincide with the rest frequencies of one the transitions: O$_3$~$[16(1,15)-16(0,16)]$, CO~(2-1), CO~(3-2), CH$_2$Cl$_2$~$[6(6,0)-5(5,1)]$~$I=1-3,F=6-7$ and  CH$_2$Cl$_2$~$[6(6,0)-5(5,1)]$~$I=1-3,F=7-8$. In particular, the 231.28 GHz absorption line was observed on two occasions: 2015.3 and 2017.9; the frequencies at  maximum absorption differed in an amount corresponding to a velocity of  about 8 km s$^{-1}$, which was  the difference in the earth velocities relative to the LSR on the two epochs.

We provide a simple explanation for the absence of absorption lines in the phase calibrator, observed almost  simultaneously with $\eta$~Car and processed with the same calibration procedure. It relies on the fact that the intensity of $\eta$~Car  (Jy beam $^{-1}$) is much higher than that of the phase calibrator (quasar) and it is also extended, while the calibrator is not resolved. The result is that the possible absorption line falls into the noise level of the calibrator but is mach larger in $\eta$~Car, where it is further increased as the spectrum is integrated along the continuum emission region. 

We conclude pointing out the importance of calibration in the interpretation of the spectra of sources with strong and spatially extended continuum emission at frequencies close to the frequencies of telluric lines, even if these lines are not as strong as O$_3$ and the baseline calibrator is observed at elevations very close to those of the target source.

\section{Data Availability}
The data underlying this article are available in  ADS/JAO.ALMA$\#2015.1.00661.S, ADS/JAO.ALMA$\#2016.1.00585.S and ADS/JAO.ALMA$\#$2017.1.00725.S

\section*{Acknowledgements}

This paper makes use of the following ALMA data: ADS/JAO.ALMA$\#2015.1.00661.S,  ADS/JAO.ALMA$\#2016.1.00585.S and ADS/JAO.ALMA$\#$2017.1.00725.S. ALMA is a partnership of ESO (representing its member states), NSF (USA) and NINS (Japan),
together with NRC (Canada), MOST and ASIAA (Taiwan), and KASI (Republic of Korea), in
cooperation with the Republic of Chile. The Joint ALMA Observatory is operated by
ESO, AUI/NRAO and NAOJ.
 We are grateful to the referee for helpful suggestion and the ALMA help desk for their assistance.
ZA and PPB acknowledge Brazilian agencies FAPESP (grant \#2011/51676-9 and \#2014/07460-0) and CNPq (grant \#304242/2019-5)




\label{lastpage}


\begin{thebibliography}{}
\bibitem[Abraham et al.(2005)]{abr05} Abraham, Z., Falceta-Gon\c calves, D., Dominici, T. et al., 2005, \aap, 437, 977
\bibitem[Abraham et al.(2020)]{abr20} Abraham, Z., Beaklini, P.P.B., Cox, P., Falceta-Gon\c calves, D., Nyman, L-\AA., \ 2020, \mnras, 499, 2493
¨
\bibitem[Bordiu \& Rizzo(2019)]{bor19} Bordiu, C., Rizzo, J. R., 2019, \mnras, 490, 1570
\bibitem[Clementel et al.(2014)]{cle14} Clementel, N., Madura, T. I., Cruip, C. J. H., Icke, V., Gull, T. R., 2014, \mnras, 443, 2475
\bibitem[Corcoran et al.(2001)]{cor01} Corcoran, M. F., Swank, J. H., Petre, R. et al., 2001, \apj, 562, 1031
\bibitem[Corcoran et al.(2017)]{cor17} Corcoran, M. F., Liburd, J., Morris, D., et al. 2017, \apj, 838, 45
\bibitem[Damineli(1996)]{dam96} Damineli, A.,  1996, \apj, 460, L49
\bibitem[Damineli et al.(2000)]{dam00} Damineli A., Kaufer A., Wolf, B.,  Stahl, O., Lopes, D.F., Ara\'ujo F. X., 2000,  \apj, 528, L101
\bibitem[Davidson \& Humphreys(1997)]{dah97} Davidson, K., Humphreys, R. M., 1997, \araa, 35, 1 
\bibitem[Davidson et al.(1997)]{dav97} Davidson, K., Ebbets, D., Johansson, S. et al., 1997, \aj, 113, 335
\bibitem[Duncan \& White(2003)]{dun03} Duncan, R. A., White, S. M., \ 2003, \mnras, 338, 425
\bibitem[Falceta-Gon\c calves, Jatenco-Pereira \& Abraham(2005)]{fal05} Falceta-Gon\c calves, D., Jatenco-Pereira, V., Abraham, Z., 2005, \mnras, 357, 895 
\bibitem[Groh et al.(2012)]{gro12} Groh, J. H., Madura, T. I., Hillier, D. J. et al., 2012, \apj, 759, L2
\bibitem[Hillier et al.(2001)]{hil01} Hillier, D. J., Davidson, K., Ishibashi, K., Gull, T., 2001, \apj, 553, 837
\bibitem[Hillier et al.(2006)]{hil06} Hillier, D. J., Gull, T., Nielsen, K. et al., 2006, \apj, 642, 1098

\bibitem[Hunter et al.(2018)]{hun18} Hunter, T., Phillips, N., Brogui\' ere, D., Gonzales, J., 2018, 
https://science.nrao.edu › alma-develop-old-022217

\bibitem[Madura et al.(2012)]{mad12} Madura, T.I., Gull, T.R., Owocki, S.P., Groh, J.H., Okazaki, A.T., Russell, C.M., 2012, \mnras, 420, 2064
\bibitem[Madura et al.(2013)]{mad13} Madura et al., 2013, \mnras, 436, 3820
\bibitem[Mehner et al.(2010)]{meh10} Mehner, A., Davidson, K., Ferland, G. J. et al.,  2010, \apj, 710, 729
\bibitem[Mehner et al. (2019)]{men19} Menher, A. and al., 2019, \aap, 630, L6
 \bibitem[Morris et al.(2017)]{mor17} Morris, P.W., Gull, T.R., Hillier, D.J. et al.,  2017, \apj, 842, 79
 \bibitem[Morris et al.(2020)]{mor20} Morris, P. W. et al., 2020, \apj, 802, 23
 \bibitem[Parkin, et al.(2011)]{par11} Parkin, E, R., Pittard, J. M., Corcoran, Hamaguchi, K.  2011, \apj, 726, 105
 \bibitem[Pittard \& Corcoran(2002)]{pit02} Pittard, J.M., Corcoran, M.F., 2002, \aap, 383, 636
 \bibitem[Reitberger et al.(2015)]{rei15} Reitberger, K., Reimer, A., Reimer, O.,Takahashi, H., \ 2015, \aap, 577, 100
 \bibitem[Russel et al.(1987)]{rus87} Russel, R. W., Lynch, D. K., Hackwell, J. A., Rudy, R. J., Rossano, G. S., 1987, \apj, 321, 937 
\bibitem[Russell et al.(2016)] {rus16} Russell, C. M., Corcoran, M. F., Hamaguchi, K. et al. 2016, \mnras, 458, 2275

\bibitem[Smette et al.(2015)]{sme15} Smette A., Sana H., Noll S., Horst H., Kausch W., Kimeswenger S., Barden M., et al., 2015, A\&A, 576, A77

\bibitem[Smith et al.(1995)]{smi95} Smith, C. H., Aitken, D. K., Moore, T. J. T., Roche, P. F., Puetter, R. C., Pi\~ na, R. K., 1995, \mnras, 273, 354
\bibitem[Smith, Ginsburg \& Bally(2018)]{smi18} Smith, N., Ginsburg, A., Bally, J.,  2018, \mnras, 474, 4988
\bibitem[Teodoro et al.(2016)]{teo16} Teodoro M., Damineli A., Heathcote B.,  Richardson N. D., Moffat A. F. J.,  2016, \apj, 819, 131
\bibitem[Verner, Bruhweiler \& Gull(2005)]{ver05} Verner, E.M., Bruhweiler, F.,  Gull, T., 2005, \apj, 624, 973 
\bibitem[Weigelt \& Ebersberger(1986)]{wei86} Weigelt, G., Ebersberger, J.,  1986, \aap, 163, L5
\bibitem[Whitelock et al.(2004)]{whi04} Whitelock, P.A., Feast, M.W.,Marang, F., Breedt, E., 2004, \mnras, 352,447
\end{thebibliography}
\end{document}